# Water Resources Research®

## RESEARCH ARTICLE

**Author Contributions:**
**Conceptualization:** Patricio Yeste, Lieke A. Melsen
**Data curation:** Matilde García-Valdecasas Ojeda
**Formal analysis:** Patricio Yeste
**Funding acquisition:** Sonia R. Gámiz-Fortis, Yolanda Castro-Díez, María Jesús Esteban-Parra

# A Pareto-Based Sensitivity Analysis and Multiobjective Calibration Approach for Integrating Streamflow and Evaporation Data

Patricio Yeste[1,2] , Lieke A. Melsen[3] , Matilde García-Valdecasas Ojeda[1,2] , Sonia R. Gámiz-Fortis[1,2] , Yolanda Castro-Díez[1,2], and María Jesús Esteban-Parra[1,2]

[1]Department of Applied Physics, University of Granada, Granada, Spain, [2]Andalusian Institute for Earth System Research (IISTA-CEAMA), University of Granada, Granada, Spain, [3]Hydrology and Quantitative Water Management Group, Wageningen University, Wageningen, The Netherlands

**Abstract** Evaporation is gaining increasing attention as a calibration and evaluation variable in hydrologic studies that seek to improve the physical realism of hydrologic models and go beyond the long-established streamflow-only calibration. However, this trend is not yet reflected in sensitivity analyses aimed at determining the relevant parameters to calibrate, where streamflow has traditionally played a leading role. On the basis of a Pareto optimization approach, we propose a framework to integrate the temporal dynamics of streamflow and evaporation into the sensitivity analysis and calibration stages of the hydrologic modeling exercise, here referred to as "Pareto-based sensitivity analysis" and "multiobjective calibration." The framework is successfully applied to a case study using the Variable Infiltration Capacity (VIC) model in three catchments located in Spain as representative of the different hydroclimatic conditions within the Iberian Peninsula. Several VIC vegetation parameters were identified as important to the performance estimates for evaporation during sensitivity analysis, and therefore were suitable candidates to improve the model representation of evaporative fluxes. Sensitivities for the streamflow performance, in turn, were mostly driven by the soil and routing parameters, with little contribution from the vegetation parameters. The multiobjective calibration experiments were carried out for the most parsimonious parameterization after a comparative analysis of the performance gains and losses for streamflow and evaporation, and yielded optimal adjustments for both hydrologic variables simultaneously. Results from this study will help to develop a better understanding of the trade-offs resulting from the joint integration of streamflow and evaporation data into modeling frameworks.

## 1. Introduction

Terrestrial water storage and water fluxes modulating the land-atmosphere interaction are key components of the global water cycle and energy budget. Water storage is in direct connection with the partitioning of precipitation into runoff and evaporation (Lehner et al., 2019; Pokhrel et al., 2021), the latter currently accounting for two-thirds of global precipitation (Good et al., 2015). The intensification of the global water cycle driven by an already changing climate represents a challenge for future water security (Lehner et al., 2019), and is likely to affect the occurrence of extreme events such as droughts and floods (Blöschl et al., 2020; Miralles et al., 2019; Peterson et al., 2021). Moreover, evaporation is projected to increase in the context of global warming (IPCC, 2021), and its changes can pose an important threat for water resources availability and the biosphere (Konapala et al., 2020; Koppa et al., 2022).

The scientific underpinning for understanding and monitoring the water cycle relies upon simulations of water stores and fluxes (Fersch et al., 2020; Pokhrel et al., 2021). Our ability to reproduce existing hydroclimatic conditions and anticipate their future changes depends on an accurate representation of hydrologic processes within both climate and hydrologic model structures (Clark et al., 2015; Koppa et al., 2022). From a hydrologic perspective, this situation brings about the need to enhance physical realism in process-based hydrologic models by developing approaches focused on multiple processes and calibration strategies that integrate multi-variate data (Clark et al., 2017, 2021; Efstratiadis & Koutsoyiannis, 2010). The incorporation of multiple data sources into the calibration stage permits to identify meaningful parameter combinations and helps achieve a more realistic representation of the hydrologic cycle (Fowler et al., 2018; Gharari et al., 2013; Koppa et al., 2019; Nijzink et al., 2018; Rakovec, Kumar, Attinger, & Samaniego, 2016; Zhang et al., 2020). This is particularly relevant for a field where streamflow has dominated calibration endeavors (Becker et al., 2019; Dembélé,

**Methodology:** Patricio Yeste, Lieke A. Melsen
**Project Administration:** Sonia R. Gámiz-Fortis, Yolanda Castro-Díez, María Jesús Esteban-Parra
**Resources:** Lieke A. Melsen
**Software:** Patricio Yeste, Matilde García-Valdecasas Ojeda
**Supervision:** Lieke A. Melsen, María Jesús Esteban-Parra
**Validation:** Patricio Yeste
**Visualization:** Lieke A. Melsen
**Writing – original draft:** Patricio Yeste
**Writing – review & editing:** Lieke A. Melsen, Sonia R. Gámiz-Fortis, Yolanda Castro-Díez, María Jesús Esteban-Parra

Ceperley, et al., 2020; Dembélé, Hrachowitz, et al., 2020) and evaporation has been considered a subproduct of the steady-state assumption for closed hydrologic systems (Han et al., 2020; Liang et al., 2015; Yang et al., 2017).

In this regard, the use of satellite-based and reanalysis evaporation data as well as in-situ measurements represents a promising solution for improving model calibration (Lettenmaier et al., 2015; Minville et al., 2014; Puertes et al., 2019; Széles et al., 2020). The growing availability of evaporation products provides an unprecedented opportunity to improve the representation of the land-atmosphere interactions within the hydrologic system, effectively overcoming the inherent limitations of the streamflow-only calibration (Dembélé, Ceperley, et al., 2020; Dembélé, Hrachowitz, et al., 2020; Koppa et al., 2019). While streamflow provides valuable information about the dynamics of the hydrologic system, evaporation data can help understand other relevant spatiotemporal processes, particularly in data-scarce areas (Dembélé, Hrachowitz, et al., 2020; López López et al., 2017) and in arid and semiarid regions where evaporative fluxes are dominant (Dembélé, Ceperley, et al., 2020; Koppa et al., 2022).

Although the use of evaporation data is becoming increasingly extended as an evaluation variable during a postcalibration phase (e.g., Bouaziz et al., 2021; Rakovec, Kumar, Attinger, & Samaniego, 2016; Rakovec et al., 2019; Rakovec, Kumar, Mai, et al., 2016; Yeste et al., 2020, 2021) or as a calibration objective itself (e.g., Dembélé, Ceperley, et al., 2020; Dembélé, Hrachowitz, et al., 2020; Demirel et al., 2018; Koppa et al., 2019; López López et al., 2017; Nijzink et al., 2018; Széles et al., 2020), evaporation-related parameters controlling the vegetation processes have remained mostly unattended during sensitivity analysis. This is particularly relevant in the light of results reported in Sepúlveda et al. (2022) for the Variable Infiltration Capacity (VIC) model (Liang et al., 1994, 1996), one of the most widely used models by the hydrologic community (Addor & Melsen, 2019), and suggest that, among others, VIC vegetation parameters have strong potential to improve model performance despite their sensitivities have been largely unknown for decades.

In this study, we propose a framework focusing on the integration of the temporal dynamics of streamflow and evaporation into the sensitivity analysis and calibration stages, which is then applied to a case study using VIC in three Spanish catchments. The hydroclimatic characteristics of these catchments (Yeste et al., 2018, 2020, 2021) together with the aridity conditions specific of the Iberian Peninsula (García-Valdecasas Ojeda et al., 2020, 2021) make them ideal candidates for the matter in question. First, we revisit the definition of the multiobjective optimization problem (MOOP) and outline how it can be addressed in practice. This will help to prepare the ground for introducing the followed approach, which is then applied to the case study. The sound performance of VIC and the broad implications derived from this study suggest a clear contribution to the body of multiple-criteria applications in hydrology, as well as to the knowledge of evaporation and its integration into the hydrologic modeling exercise.

## 2. Multiobjective Optimization Problem

Multiobjective optimization aims at optimizing multiple objectives at the same time. Without loss of generality, the MOOP is commonly formulated as a minimization problem (Miettinen, 1998), and can be formally written as follows (Deb, 2001):

$$
\left.
\begin{array}{lll}
\text{Minimize} & f_m(\mathbf{x}) & m = 1, 2, \ldots, M \\
\text{subject to} & g_j(\mathbf{x}) \leq 0 & j = 1, 2, \ldots, J \\
& h_k(\mathbf{x}) = 0 & k = 1, 2, \ldots, K \\
& x_i \in \left[ x_i^{(L)}, x_i^{(U)} \right] & i = 1, 2, \ldots, N
\end{array}
\right\}
\tag{1}
$$

where $M$ is the number of objective functions $f_m$ to minimize. Each component $x_i$ of the $n$-dimensional decision vector $\mathbf{x} = (x_1, x_2, \ldots, x_N)$ must belong to the interval defined by the corresponding lower and upper bounds $x_i^{(L)}$ and $x_i^{(U)}$. The MOOP can be subject to $J$ inequality and $K$ equality constraints applied to the constraint functions $g_j$ and $h_k$, respectively.

By definition, a conflict between the objectives leads to a trade-off in Pareto-optimal solutions that are equally important or likely from an optimization perspective, conforming the so-called Pareto front (Deb, 2001). As none of the solutions can be considered superior when there is more than one objective to minimize, Pareto-optimal solutions are also referred to as nondominated solutions. The MOOP then becomes a multicriteria decision-making problem where only a small subset of solutions is selected for the specific application based on expert knowledge or the use of decision-making techniques (Miettinen, 1998).

Classical approaches to solve an MOOP rely on the scalarization of the multiple objectives into a composite single-objective function. The most common form of scalarization involves the introduction of some preference information for the objectives through weighting coefficients, which can be changed in order to obtain alternative Pareto-optimal solutions (Deb, 2001; Miettinen, 1998). Among these techniques, the weighted metric method represents the most general problem for minimizing distances (Deb, 2001)

$$\text{Minimize} \quad \left( \sum_{m=1}^{M} w_m |f_m(\mathbf{x}) - z_m^*|^p \right)^{1/p} \tag{2}$$

where $w_m$ is the weight for the objective function $f_m$, $z_m$ is the $m$th component of the ideal vector $z^*$ (usually the null vector) and $p$ is a power parameter. The weights are typically selected such that $w_m \geq 0$ and their sum is one. The problem can be subject to the same constraints as in Equation 1, and the parameter $p$ can take values between 1 and $\infty$. Notably, when $p = 2$ the problem corresponds to minimizing the weighted Euclidean distance to the ideal solution $z^*$. In the particular case of $p = 2$ and $M$ equal weights, the problem is equivalent to minimizing the pure Euclidean distance.

A major disadvantage of classical methods is that only one solution is handled at a time (Deb, 2001). In the event of nonconvex and discontinuous Pareto fronts, this unavoidably results in an incomplete representation of the optimal solutions region (Miettinen, 1998). In classical methods relying on weights, for instance, there can be regions in the Pareto front that remain unreachable for any combination of weights due to the existence of nonconvexities and discontinuities (Deb, 2001; Miettinen, 1998). By contrast, multiobjective optimization evolutionary algorithms (MOEAs) follow a metaheuristic approach that effectively handles a population of candidate solutions in a single run, making them suitable for nonlinear optimization problems regardless of any continuity or convexity condition (Coello Coello, 2005; Deb, 2001). MOEAs are then left with the task of converging to the Pareto front and achieving a well-distributed set of solutions (Abraham & Jain, 2005; Deb, 2001; Laumanns, 2005; Yang et al., 2014). Metaheuristic approaches have become the way forward to tackle multiobjective optimization challenges in many areas of science and engineering, and the efforts to improve and develop them continue to the present day (Blank & Deb, 2020; Falcón-Cardona et al., 2021; Hernandez-Suarez et al., 2021; LaTorre et al., 2021).

## 3. Approach Definition

In this section, the methodical steps toward our goal of integrating streamflow and evaporation data into the model optimization process are presented. These steps are schematized in Figure 1, and their description is provided in the following subsections. The core idea underlying the rationale of this framework consists in the application of the MOOP definition to two objectives that characterize the overall goodness-of-fit of the model regarding the temporal dynamics of the streamflow and evaporation data, respectively. In particular, we propose a multiobjective optimization approach that can serve the purpose of both a sensitivity analysis and a calibration procedure that are performed in tandem.

But first, it is essential to verify if the problem is affordable from a physical perspective. As we will show below, the integration of both data sources into the modeling framework is largely dependent on the law of conservation of mass and its steady-state representation for the hydrologic system under examination.

### 3.1. Verification for Streamflow and Evaporation Data

Over long periods, the water balance equation can be written as follows after normalizing by precipitation (Han et al., 2020; Mendoza et al., 2015, 2016; Rasmussen et al., 2014; Yeste et al., 2021)

$$1 = \frac{\overline{Q}}{\overline{P}} + \frac{\overline{E}}{\overline{P}} \tag{3}$$

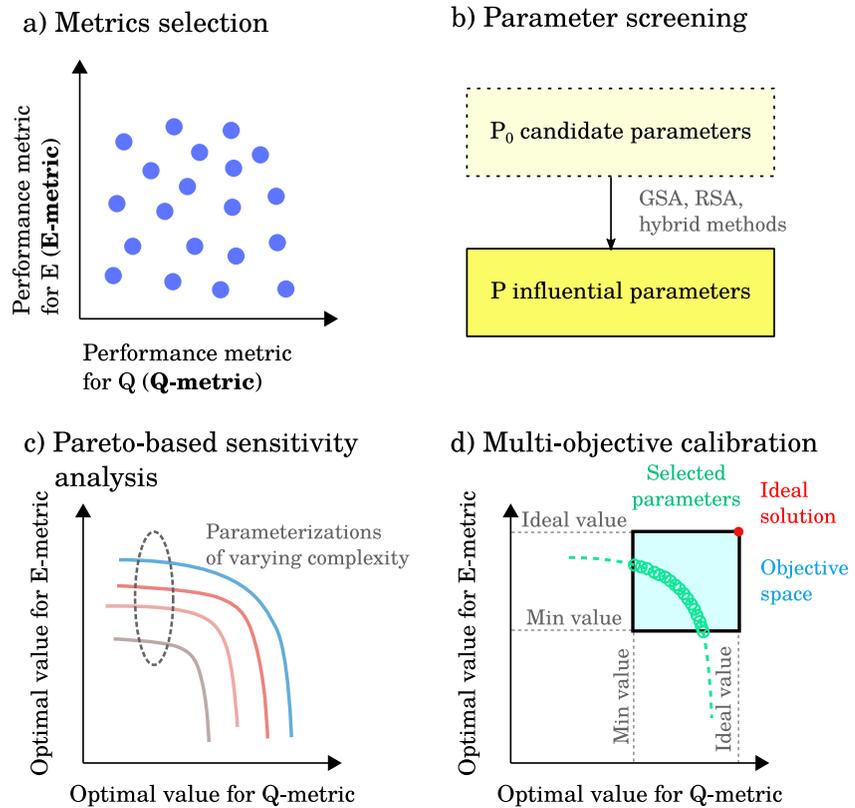

**Figure 1.** Schematic overview of the framework. (a) Metrics selection for streamflow and evaporation based on model performance. (b) Parameter screening to identify influential parameters. (c) Pareto-based sensitivity analysis for parameterizations of varying complexity. (d) Multiobjective calibration for the final selection of parameters.

where $\overline{P}$ $[L/T]$ is the mean annual precipitation for the period, $\overline{Q}$ $[L/T]$ the mean annual streamflow, and $\overline{E}$ $[L/T]$ is the mean annual evaporation.

Equation 3 shows that different model simulations forced with same input $P$ data can reach the steady-state for different contribution levels of $Q$ and $E$. Hence, for a long enough period, an imbalance in the above equation when calculated for the reference data constitutes a potential condition for a trade-off between the objectives (i.e., a $Q$-$E$ trade-off) since the model will likely produce a steady-state solution. If the imbalance is significant, the integration of the streamflow and evaporation data sets could become unsuitable due to a failure to comply with the law of conservation of mass. For small to negligible imbalances, however, there is no a priori condition for a $Q$-$E$ trade-off. In that case, the trade-off will generally depend on the particular application and the (in)ability of the model to simultaneously reproduce both data sets. Equation 3 will be thoroughly checked for the case study.

### 3.2. Metrics Selection and Parameter Screening

The *metrics selection* step (Figure 1a) is based on a preliminary evaluation of model performance by means of several candidate performance metrics for $Q$ and $E$. These metrics are expressed as single scalar values aggregating different statistical properties of model performance (e.g., Nash-Sutcliffe Efficiency (NSE), Kling-Gupta Efficiency (KGE), Mean Squared Error), and can be calculated at different time scales (e.g., daily, monthly). This stage requires a large-enough and representative sample of model responses, which can be achieved via the sampling method of the selected sensitivity analysis method during the following step in order to leverage the simulated responses and guarantee computational efficiency. As a result, a Q-metric and an E-metric are selected from the proposed candidates according to the maximum performance attainable for streamflow and evaporation. Next, a *parameter screening* procedure (Figure 1b) is carried out to identify influential and noninfluential parameters to the selected metrics. Sensitivity analysis methods play a major role in this respect as they are traditionally used to this end. In this way, global sensitivity analysis (GSA), regional sensitivity analysis (RSA), and hybrid local-global methods are equally valid approaches that can be conducted for that purpose.

### 3.3. Pareto-Based Sensitivity Analysis and Multiobjective Calibration

The last two steps are founded on an MOOP implementation for the selected metrics. In the first place, the parameter screening procedure is further refined in a Pareto setting. We call this step *Pareto-based sensitivity analysis* (Figure 1c) since it relies on the MOOP definition and is a continuation of the previously applied sensitivity analysis that aims at finding the most parsimonious parameterization to satisfactory results. This is attained through unconstrained multiobjective optimization experiments targeted at parameterizations of varying complexity as follows: starting from a reference parameterization that covers the minimum number of parameters to be analyzed, the rest of the parameters are included one at a time. All the parameters are then incorporated into a full parameterization that reflects the maximum achievable complexity for the given implementation. The Pareto front of a parameterization including one additional parameter is indicative of the sensitivity for that particular parameter when compared to the front of the reference parameterization. On the other hand, the Pareto front corresponding to the full parameterization points to a higher-order (i.e., interaction) effect. In this work, the interaction effects between parameters are exclusively assessed on the basis of the full parameterization, but the analysis can be easily extended for combinations of two or more extra parameters covering the range of complexity between the reference and the full parameterization. This stage ends with the selection of one of the parameterizations based on the relative gain/loss in performance for both objectives. Optimal model performance is finally determined for the selected parameterization through a constrained *multiobjective calibration* exercise (Figure 1d) focused on a specific region of the objective space in order to discard unfeasible solutions leading to poor performance estimates.

## 4. Case Study

This section begins with the presentation of the study catchments and data and the hydrologic model that integrate the case study, and is then followed by a description of how the four steps constituting our framework are implemented for the three catchments.

### 4.1. Area and Data

In order to assess the applicability of the proposed framework, the approach was implemented for three catchments in Spain corresponding to the following stations (Figure 2): station R-5055 "José Torán," R-5005 "Rumblar," and R-2011 "Arlanzón." Stations R-5055 and R-5005 belong to the Guadalquivir River Basin, the southern-most, and one of the most arid basins in Spain, while station R-2011 is located in the Duero River Basin, a transboundary basin that flows into northern Portugal, and the biggest basin of the Iberian Peninsula. The Guadalquivir and Duero River Basins perfectly represent the latitudinal gradient of the climate variability for the Iberian Peninsula, and therefore are appropriate candidates to test the spatial consistency of the applied methods. We refer to Yeste et al. (2018) and Yeste et al. (2020, 2021) and references therein for a comprehensive hydroclimatic characterization of these basins.

Hydrologic data in this study were assembled for a period of 10 hydrologic years spanning from October 2000 to September 2010, which was chosen as the study period in this work. Daily precipitation and temperature observations were extracted from the Spanish PREcipitation At Daily scale (SPREAD, Serrano-Notivoli et al., 2017) and the Spanish TEmperature At Daily scale (STEAD, Serrano-Notivoli et al., 2019) data sets, respectively. SPREAD and STEAD are high-resolution (5 km) gridded data sets for the Spanish domain that integrate meteorological information from a dense network of observatories. Daily streamflow time series for the studied catchments were acquired from the Spanish Center for Public Work Experimentation (CEDEX), and evaporation data were gathered from the daily outputs of the Global Land Evaporation Amsterdam Model (GLEAM) version 3.5a (Martens et al., 2017; Miralles et al., 2011) at 0.25° resolution. As in Yeste et al. (2020), additional meteorological forcings (i.e., wind speed, radiation, atmospheric pressure, vapor pressure), were collated from high-resolution (0.088°) simulations conducted with the Weather Research and Forecasting (WRF) using ERA-Interim Reanalysis data in García-Valdecasas Ojeda et al. (2017).

The ratios of streamflow and evaporation to precipitation ($\overline{Q/P}$ and $\overline{E/P}$) corresponding to the study period were computed for all the catchments and produced the following estimates: 0.33 and 0.67 for R-5055; 0.24 and 0.84 for R-5005, and 0.56 and 0.41 for R-2011. These values do not introduce an artificial imbalance as their sum remains close to 1, and therefore guarantee the further applicability of the framework (see Section 3.1). The high values of $\overline{E}/\overline{P}$ for the three catchments are symptomatic of the aridity conditions of the Guadalquivir and Duero River Basins, and thus reinforce the potential of including evaporation information into the modeling framework (Dembélé, Ceperley, et al., 2020; Koppa et al., 2022).

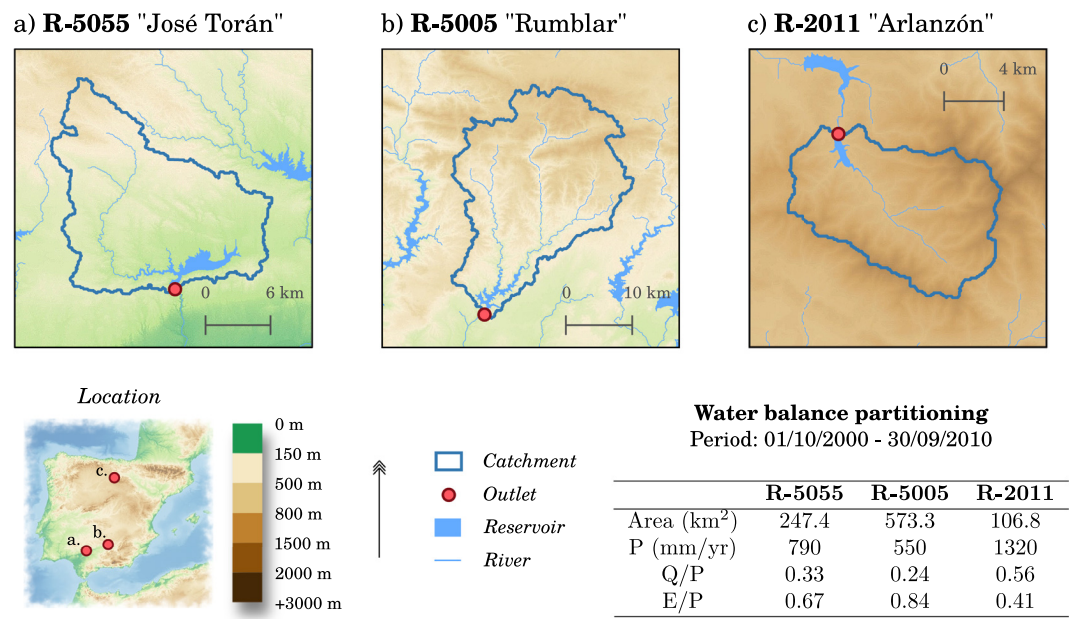

**Figure 2.** Location of the three studied catchments. (a) Catchment R-5055 "José Torán." (b) Catchment R-5005 "Rumblar." (c) Catchment R-2011 "Arlanzón."

### 4.2. Hydrologic Model

#### 4.2.1. The VIC Model

The VIC (Liang et al., 1994, 1996) model (version 4.2.d) is a semidistributed macroscale hydrologic model that implements the water and energy balance equations for a gridded domain at daily and subdaily time steps. The subgrid variability in land-uses is handled statistically through vegetation tiles for which both equations are solved. Evaporation for every grid cell is limited by the atmospheric demand of water vapor, or potential evaporation, according to the Penman-Monteith equation, and is calculated as the sum of three components: evaporation from interception, transpiration, and evaporation from bare soil. The soil profile of each grid cell is divided into three layers in order to conceptualize the runoff generation process. Surface runoff is generated from the first two soil layers following the Xinanjiang formulation (Zhao, Ren-Jun, et al., 1980), and baseflow is generated in the bottom layer via an Arno equation (Franchini & Pacciani, 1991). The model was implemented with a gridded configuration at 0.05° resolution (see Yeste et al., 2020) and a spin-up period of 20 years prior to the study period. The meteorological forcings and the evaporation data were interpolated to match the resolution of the VIC model using a nearest neighbor assignment.

The VIC model does not consider the horizontal fluxes between neighboring grid cells. This is often accounted for by coupling a routing scheme in a postprocessing phase. In this study, a gamma function with two parameters (shape and scale) was chosen as the unit hydrograph representing the time delay between the runoff generated within the catchment and the streamflow draining through the catchment outlet. The simulated runoff was weighted for each grid cell by the fractional area falling inside the catchment. This approach adequately respects the real shape of the catchment, and is a common choice for hydrologic studies using VIC (e.g., Mizukami et al., 2017; Rakovec et al., 2014, 2019).

#### 4.2.2. Parameter Selection

First, the soil and vegetation parameters considered observable were gathered from the following data sources: (a) bulk density and soil textural classes were extracted from SoilGrids1km (Hengl et al., 2014); (b) porosity, saturated hydraulic conductivity, field capacity, and wilting point were collected from EU-SoilHydroGrids

**Table 1**
*Parameters Included in the Parameter Screening Procedure*

| Parameter | Units | Min value | Max value | Description |
|---|---|---|---|---|
| $b_i$ | – | $10^{-5}$ | 0.4 | Variable infiltration shape parameter |
| $D_S$ | – | $10^{-9}$ | 1 | Fraction of $D_m$ where nonlinear baseflow begins |
| $W_S$ | – | $10^{-9}$ | 1 | Fraction of the porosity of the bottom soil layer where nonlinear baseflow begins |
| $D_m$ | mm/day | $10^{-9}$ | 30 | Maximum baseflow |
| $d_2$ | m | 0.1 | 0.9 | Thickness of soil layer 2 |
| $rout_1$ | – | $10^{-9}$ | 10 | Shape parameter of gamma function |
| $rout_2$ | days | $10^{-9}$ | 2 | Scale parameter of gamma function |
| $depth_{1f}$ | – | 0.5 | 1.5 | Thickness of root zone for layer 1 |
| $depth_{2f}$ | – | 0.5 | 1.5 | Thickness of root zone for layer 2 |
| $rarc_f$ | – | 0.5 | 1.5 | Architectural resistance |
| $rmin_f$ | – | 0.5 | 1.5 | Minimum stomatal resistance |
| $LAI_f$ | – | 0.5 | 1.5 | Leaf-area index (12 monthly values) |
| $albedo_f$ | – | 0.5 | 1.5 | Albedo (12 monthly values) |
| $rough_f$ | – | 0.5 | 1.5 | Vegetation roughness (12 monthly values) |
| $disp_f$ | – | 0.5 | 1.5 | Vegetation displacement (12 monthly values) |
| $wind\_h_f$ | – | 0.5 | 1.5 | Height of wind speed measures |
| $RGL_f$ | – | 0.5 | 1.5 | Minimum incoming shortwave radiation for transpiration |
| $rad\_atten_f$ | – | 0.5 | 1.5 | Radiation attenuation |
| $wind\_atten_f$ | – | 0.5 | 1.5 | Wind speed attenuation through overstory |
| $trunk\_ratio_f$ | – | 0.5 | 1.5 | Ratio of total tree height that is trunk |

*Note.* Parameters denoted with subscript "$f$" correspond to the dimensionless multiplication factors used to modify VIC vegetation parameters.

ver1.0 (Tóth et al., 2017); (c) land-uses were adopted from the UMD Global Land Cover Classification (Hansen et al., 2000), and the associated vegetation parameters were defined in agreement with the Global Land Data Assimilation System (GLDAS) specifications for VIC (Rodell et al., 2004). All the soil and vegetation data sets were collated at 1-km resolution and were regridded to the model resolution using a first-order conservative remapping to preserve the integral of the values.

A set of 20 parameters representing the soil, vegetation, and routing processes was subsequently selected for the parameter screening procedure, and comprised 5 soil parameters, the 2 routing parameters, and 13 vegetation parameters (Table 1). Although the soil and routing parameters considered here are commonly used in other studies for sensitivity analysis and/or calibration purposes (e.g., Chawla & Mujumdar, 2015; Mizukami et al., 2017; Oubeidillah et al., 2014; Rakovec et al., 2014, 2019; Yeste et al., 2020), the vegetation parameters remain mostly fixed as default values and are rarely varied during the hydrologic modeling exercise (see Melsen et al., 2016; Mendoza et al., 2015; Sepúlveda et al., 2022, for some exceptions).

The values of the vegetation parameters were modified by means of multiplication factors varying from 0.5 to 1.5. Multiplication factors confer greater flexibility on the VIC parameterization of the vegetation processes, and were applied to both the single-valued parameters and those with monthly variations (see Table 1). This strategy is comparable to the use of spatial multipliers to calibrate distributed hydrologic models (e.g., Bandaragoda et al., 2004; Francés et al., 2007; Mendoza et al., 2015; Pokhrel & Gupta, 2010), and allows for studying the potential of the VIC vegetation parameters as calibration targets (Sepúlveda et al., 2022).

### 4.3. Performance Metrics

In this work, the NSE (Nash & Sutcliffe, 1970) was chosen to evaluate the goodness-of-fit of the model, and the metrics for streamflow and evaporation were selected according to the relative performance of NSE at daily and monthly time scales (Figure 1a). NSE can be expressed as

$$\text{NSE} = 1 - \frac{\sum_{t=1}^{T} (y_t - y_{\text{ref},t})^2}{\sum_{t=1}^{T} (y_{\text{ref},t} - \bar{y}_{\text{ref}})^2} \tag{4}$$

where $y_t$ is the model output and $y_{\text{ref},t}$ is the reference value of variable $y$ for time step $t$, and $\bar{y}_{\text{ref}}$ is the mean value of reference data for the study period. NSE is a multiple-criteria estimator of model performance, and can also be formulated as an algebraic decomposition into three components representing different statistical properties of model behavior (Clark et al., 2021; Gupta et al., 2009; Knoben et al., 2019; Lamontagne et al., 2020).

$$\text{NSE} = 2\alpha r - \alpha^2 - \frac{(\beta - 1)^2}{\text{CV}_{\text{ref}}^2} \tag{5}$$

here, $r$ is the correlation coefficient between simulations and reference data, $\alpha$ is the ratio of the standard deviations, $\beta$ is the ratio for mean values, and $\text{CV}_{\text{ref}}$ is the coefficient of variation of the reference data. $\alpha$ and $\beta$ report, respectively, on the overestimation ($>1$) or underestimation ($<1$) of the variability and the mean value of the analyzed variable. $\text{CV}_{\text{ref}}$ is model-independent, and consequently NSE can be seen as an interplay between $r$, $\alpha$, and $\beta$ (see Gupta et al. (2009) and Knoben et al. (2019) for a detailed description of these statistics and their relationships).

The $\beta$ component (i.e., the bias component) for streamflow and for evaporation are defined as $\beta_Q = \overline{Q}/\overline{Q}_{\text{ref}}$ and $\beta_E = \overline{E}/\overline{E}_{\text{ref}}$, respectively, where $\overline{Q}_{\text{ref}}$ and $\overline{E}_{\text{ref}}$ are the mean annual values of the $Q$ and $E$ reference data, respectively. Two model states $\left(\overline{Q}_1, \overline{E}_1\right)$ and $\left(\overline{Q}_2, \overline{E}_2\right)$ satisfying Equation 3 will produce the following equality:

$$\overline{Q}_1 + \overline{E}_1 = \overline{Q}_2 + \overline{E}_2 = \overline{P} \Rightarrow \overline{Q}_2 - \overline{Q}_1 = -\left(\overline{E}_2 - \overline{E}_1\right) \Rightarrow \Delta\overline{Q} = -\Delta\overline{E} \tag{6}$$

Then, from the definition of $\beta_Q$ and $\beta_E$, two model states fulfilling the above relationship will also satisfy the following equality for the differences of $\beta_Q$ and $\beta_E$:

$$\frac{\Delta\beta_E}{\Delta\beta_Q} = \frac{\Delta\overline{E}/\overline{E}_{\text{ref}}}{\Delta\overline{Q}/\overline{Q}_{\text{ref}}} = -\frac{\overline{Q}_{\text{ref}}}{\overline{E}_{\text{ref}}} \tag{7}$$

Equation 7 indicates that simulations achieving the steady-state will manifest a data-driven trade-off for $\beta_Q$ and $\beta_E$ as a straight line with negative slope that only depends on reference data. The implications of selecting NSE as a performance metric and its decomposition into $r$, $\alpha$ and $\beta$ will be discussed in the light of results.

### 4.4. Monte Carlo Experiment and DELSA Sensitivity Analysis

A Monte Carlo experiment was conducted for each of the studied catchments using a Latin Hypercube sample (LH sample) of 10,000 parameter combinations for the 20 parameters listed in Table 1. NSE was calculated at daily and monthly time scales for the streamflow and evaporation simulations, and the best performing metric in each case was selected based on a comparative study. Parameter sensitivities with respect to the two selected metrics were studied by applying the Distributed Evaluation of Local Sensitivity Analysis (DELSA) method (Rakovec et al., 2014), and the most influential parameters were selected for each catchment.

DELSA is a hybrid local-global method that estimates local gradients at several locations across parameter space. Each parameter is individually perturbed by a fixed amount at every location (usually 1%, see Melsen & Guse, 2019; Melsen et al., 2016; Rakovec et al., 2014; Sepúlveda et al., 2022) so as to estimate the first-order partial derivatives, which are an indicator of sensitivity. In this study, 1% perturbations were carried out for a population of 500 points (hereafter referred to as DELSA sample) that were extracted from the LH sample based on their proximity to the ideal case of NSE = 1 for both streamflow and evaporation. This initialization strategy is convenient insofar as it allows for estimating parameter sensitivities in the region of interest from an optimization perspective, thus making additional use of the outputs from the Monte Carlo experiment.

### 4.5. Multiobjective Optimization Strategy and Single-Objective Experiments

In terms of eligibility, the approach followed in this study is open to any multiobjective optimization algorithm and customization strategy one may apply. In our case study, we selected two optimization algorithms: the Non-Dominated Sorted Genetic Algorithm II (NSGA-II, Deb et al., 2002) and the Multiobjective Flower Pollination Algorithm (MOFPA, Yang et al., 2014). NSGA-II is a well-established evolutionary algorithm that has been extensively used in many optimization endeavors given its fast and efficient handling of an evolving population of solutions (Ercan & Goodall, 2016). MOFPA is a nature-inspired algorithm that mimics self-pollination effects of flowering plants and cross-pollination processes over long distances carried out by pollinators obeying a Lévy flight behavior (Pavlyukevich, 2007; Yang, 2012; Yang & Deb, 2010). The combination of two or more algorithms to approximate the Pareto front attempts to overcome the inability of one single optimizer to invariably perform well regardless of the problem to be solved (also known as the "No Free Lunch" theorem, see Wolpert & Macready, 1997), and constitutes an effective measure to ameliorate this problem (Faris et al., 2016; Sun et al., 2020; Vrugt et al., 2009) and a suitable method to explore complex objective spaces (Maier et al., 2014; Vrugt & Robinson, 2007).

On the other hand, the nature of such an approach implies that there is a fine line between the Pareto-based sensitivity analysis and the multiobjective calibration, it being the use (or not) of constraints. Constraints applied during the multiobjective calibration exercise were handled through a parameterless penalty approach (Deb, 2000), and were defined according to the results obtained in the prior step. Despite being commonly ignored in multicriteria applications, constraints remain a key aspect of optimization (Blank & Deb, 2020). Indeed, setting limits of acceptability for the objectives ensures that optimal solutions cover exclusively the region of interest of the objective space, and has shown to improve the streamflow representation under changing climate conditions (Fowler et al., 2018) as well as the possibility to effectively incorporate different water fluxes in the calibration framework (Koppa et al., 2019).

Finally, five single-objective experiments were conducted for station R-5055 using the Shuffled Complex Evolution (SCE-UA) algorithm (Duan et al., 1994) in order to benchmark the relative performance of different calibration strategies. These experiments were based on the weighted metric method (Equation 2) and included: a streamflow-only calibration, an evaporation-only calibration, and three weighted Euclidean distance experiments. Table 2 summarizes the different calibration strategies evaluated, and the implementational details of the employed algorithms are provided in Supporting Information S1.

## 5. Results

For the sake of clarity, the results for station R-5055 have been chosen as the guiding thread facilitating the presentation of this section. Figures corresponding to stations R-5005 and R-2011 are collected in Supporting Information S1, and will be appropriately referred to throughout the section.

### 5.1. Metrics Selection and DELSA Sensitivity Measures

First, a Monte Carlo experiment was conducted to explore parameter space for the LH sample (Figure 3). The metrics selection was carried out based on the relative performance of VIC at daily and monthly time scales both for streamflow (Figure 3a) and evaporation (Figure 3b). The model yielded daily NSE($Q$) (NSE($Q_d$) hereinafter) values above 0.6 (Figure 3a), suggesting that NSE($Q_d$) was an appropriate metric to evaluate the streamflow performance. On the contrary, NSE($E$) reflected poorer results at daily scale, with monthly estimates (NSE($E_m$)

**Table 2**
*Multiobjective and Single-Objective Calibration Experiments*

| Experiment | Number of objectives | Objective(s) to minimize |
|---|---|---|
| Multiobjective calibration | 2 | $1 - NSE(Q_d)$, $1 - NSE(E_m)$ |
| Evaporation-only calibration | 1 | $1 - NSE(E_m)$ |
| Weighted Euclidean distance | 1 | $\sqrt{0.75 \cdot (1 - NSE(E_m))^2 + 0.25 \cdot (1 - NSE(Q_d))^2}$ |
| Weighted Euclidean distance | 1 | $\sqrt{0.50 \cdot (1 - NSE(E_m))^2 + 0.50 \cdot (1 - NSE(Q_d))^2}$ |
| Weighted Euclidean distance | 1 | $\sqrt{0.25 \cdot (1 - NSE(E_m))^2 + 0.75 \cdot (1 - NSE(Q_d))^2}$ |
| Streamflow-only calibration | 1 | $1 - NSE(Q_d)$ |

*Note.* Single-objective experiments were implemented only for station R-5055.

hereinafter) always above their daily counterparts (Figure 3b). In this case, $NSE(E_m)$ was chosen as the metric of interest, which is also in line with GLEAM not being an observational data set. $NSE(Q_d)$ and $NSE(E_m)$ were also selected for the other two catchments (Figures S1 and S7 in Supporting Information S1).

The DELSA sensitivity analysis method was subsequently applied to study how model parameters affected the selected metrics. The DELSA sample was obtained from the joint representation of both metrics using a radial filter over the LH sample in order to extract the 500 points closest to the top-right corner of the scatterplot (Figure 3c). This was done to investigate sensitivity only in the region representing the highest model performance. As shown in Figure 3d, the 1% perturbations for the DELSA sample led to NSE scores close to those corresponding to the original 500 points, thus ensuring a suitable coverage of this area for the parameter screening procedure.

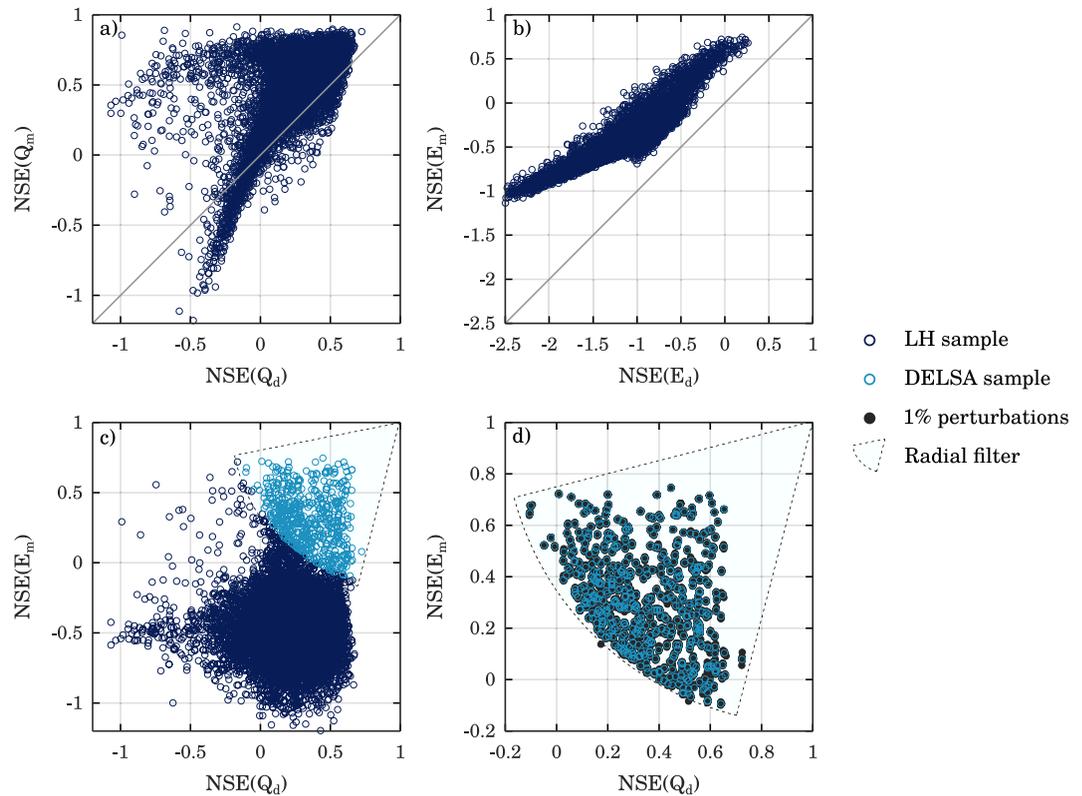

**Figure 3.** Results of the Monte Carlo experiment for station R-5055 to study parameter space. (a) Daily and monthly estimates of Nash-Sutcliffe Efficiency (NSE) for streamflow ($NSE(Q_d)$ versus $NSE(Q_m)$) corresponding to the Latin Hypercube sample (LH sample). (b) Daily and monthly estimates of NSE for evaporation ($NSE(E_d)$ versus $NSE(E_m)$) corresponding to the LH sample. (c) Joint representation of the selected performance metrics, $NSE(Q_d)$ and $NSE(E_m)$, and radial filter to extract the closest 500 points to the ideal solution (Distributed Evaluation of Local Sensitivity Analysis (DELSA) sample). (d) DELSA sample and the 1% perturbations where DELSA sensitivity analysis is to be performed.

The DELSA first-order sensitivity measures for streamflow and evaporation are depicted in Figure 4. The local DELSA measures range from 0 to 1, with values closer to 1 indicating a higher parameter sensitivity, and are commonly averaged over the complete parameter space so as to quantify and compare the individual effect of the different parameters (e.g., Melsen & Guse, 2019; Melsen et al., 2016; Rakovec et al., 2014). In order to facilitate the parameter screening procedure, Figure 4 includes the moving average of the DELSA first-order sensitivity indices for each parameter. This is a convenient way to compare the relative influence of the candidate parameters that also reflects the variability of parameter sensitivities across the objective space (i.e., the space of NSE($Q_d$) and NSE($E_m$) estimates).

Eleven out of the 20 parameters evaluated with DELSA were identified as important with respect to at least one of the performance metrics. This included the five soil parameters ($b_i$, $D_S$, $W_S$, $D_m$, $d_2$), the two routing parameters ($rout_1$, $rout_2$) and four vegetation parameters ($depth_{1f}$, $depth_{2f}$, $rmin_f$, $LAI_f$). Among soil parameters, $D_S$ and $D_m$ evoked the strongest model response in terms of NSE($Q_d$) and NSE($E_m$) (Figures 4a and 4b). The two routing parameters were the most important parameters to NSE($Q_d$), while the absence of a feedback mechanism between the VIC model structure and the routing scheme implied a null effect on NSE($E_m$). The four vegetation parameters only affected the sensitivity measures for NSE($E_m$), although their effects were substantially lower compared to $D_S$ and $D_m$. As for the other catchments, the soil and routing parameters produced similar sensitivities to that obtained for R-5055, while the number of important vegetation parameters varied from two for R-5005 (Figure S2 in Supporting Information S1) to five for R-2011 (Figure S8 in Supporting Information S1).

## 5.2. Pareto-Based Sensitivity Analysis

As shown in Figure 5a, a total of six different parameterizations were evaluated during this stage by progressively increasing the degrees of freedom of VIC when optimized in a Pareto setting: first, a reference parameterization was defined as the combination of the five soil parameters and the two routing parameters (SR parameterization); second, the SR parameterization was further extended by including each of the four vegetation parameters one at a time (SR + $depth_{1f}$, SR + $depth_{2f}$, SR + $rmin_f$ and SR + $LAI_f$); finally, the full parameterization encompassed all the 11 parameters identified with DELSA. An analogous approach was followed during the Pareto-based sensitivity analysis for the other two catchments (Figures S3a and S9a in Supporting Information S1).

The Pareto fronts obtained for the SR parameterization and the full parameterization represent the low-end and top-end performances that can be attained by VIC for each individual implementation. The greatest increase in model performance when including all the parameters was achieved for NSE($E_m$) in all the catchments, while the NSE($Q_d$) improvement was less pronounced. This is supported by the results of DELSA, as the SR parameterization already contains the most important parameters to NSE($Q_d$) (Figure 4a, Figures S2a and S8a in Supporting Information S1).

Other Pareto fronts demonstrate the individual effects of the vegetation parameters added to the SR parameterization. It is visible for R-5055 that the addition of the $LAI_f$ parameter produced a substantial increase in performance for NSE($E_m$) close to the full parameterization. The SR + $LAI_f$ parameterization was consequently adopted during the multiobjective calibration exercise (Figure 5b) for it exhibited the biggest performance enhancement and only required one extra degree of freedom with respect to the SR parameterization. Similarly, both $LAI_f$ and $rmin_f$ were added to the SR parameterization for R-5005 (Figure S3b in Supporting Information S1), and $depth_{1f}$ and $depth_{2f}$ in the case of R-2011 (Figure S9b in Supporting Information S1).

## 5.3. Multiobjective Calibration and Predictive Uncertainty

The calibration of the VIC model was carried out through a constrained Pareto optimization based on the SR + $LAI_f$ parameterization (Figure 5b). The limits of acceptability for both NSE($Q_d$) and NSE($E_m$) were set at 0.6 based on the unconstrained optimization results. The predictive intervals for the constrained Pareto optimization showed a good coverage of the streamflow observations and GLEAM data (Figure 6), highlighting the capability of VIC to reproduce both variables simultaneously. The constrained Pareto optimization for R-5005 was done subject to NSE values greater than 0.5 (Figure S3b in Supporting Information S1), and the solutions also reflect a good adjustment to streamflow and evaporation data (Figure S4 in Supporting Information S1). In the case of R-2011, the constraints were set at 0.6, resulting in a less pronounced *Q-E* trade-off (Figure S9b in Supporting Information S1) and a narrower predictive interval (Figure S10 in Supporting Information S1).

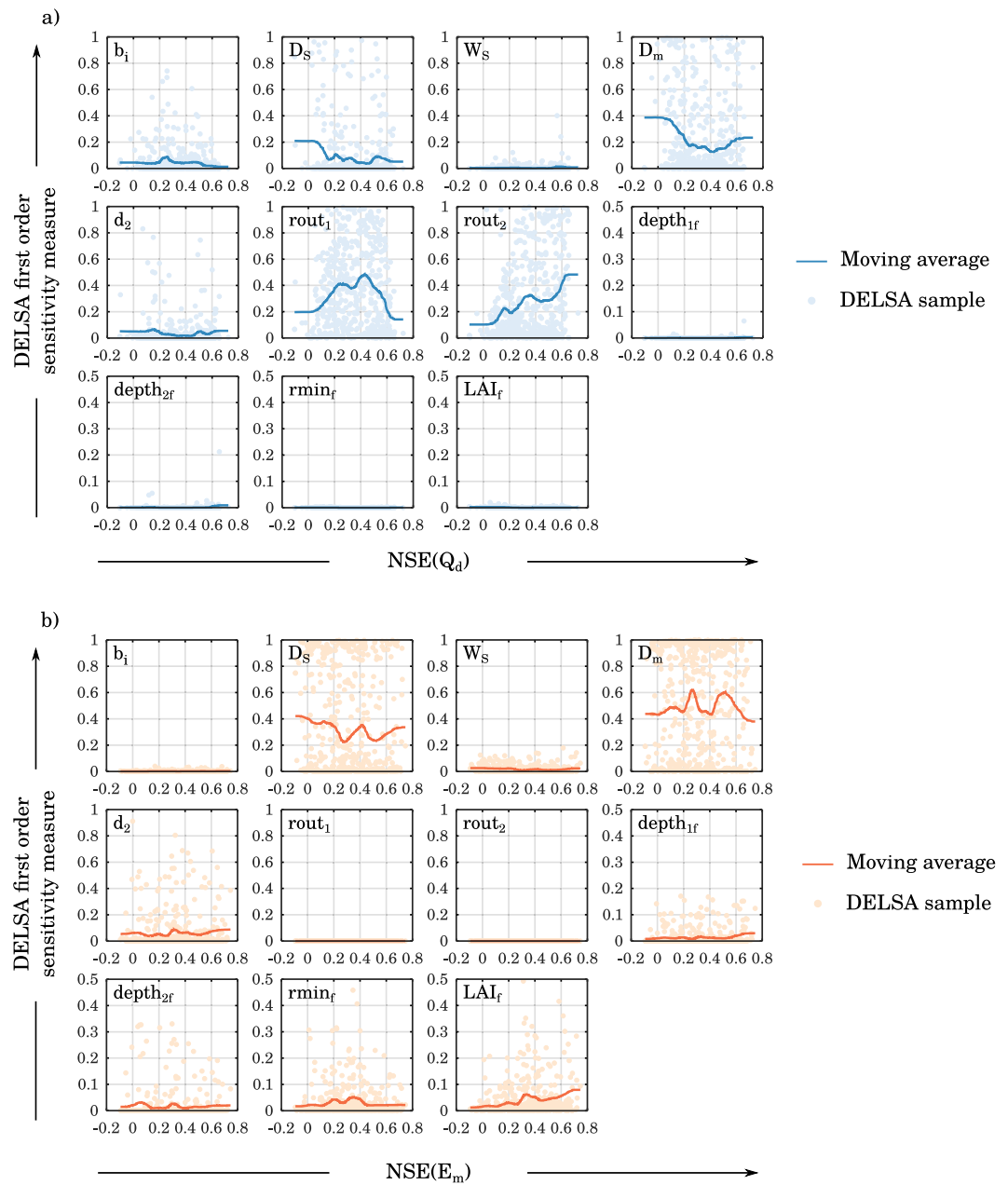

**Figure 4.** Distributed Evaluation of Local Sensitivity Analysis (DELSA) first-order sensitivities against model performance for station R-5055 computed for streamflow (a) and evaporation (b). Only the most influential parameters for at least one of the selected metrics are shown. The moving average has been depicted on top of the DELSA sample in order to facilitate the screening procedure.

To provide a more complete picture of the multiobjective calibration results, the Pareto front obtained was divided into three clusters representing different model behaviors for which the parameter distributions were compared (Figure 7): cluster C1 represents the region of the front where $NSE(E_m)$ is maximum and $NSE(Q_d)$ improves; cluster C2 is characterized by varying $NSE(E_m)$ values while $NSE(Q_d)$ remains almost constant; finally, cluster C3 includes points where $NSE(Q_d)$ is maximum and both $NSE(Q_d)$ and $NSE(E_m)$ are slightly changing.

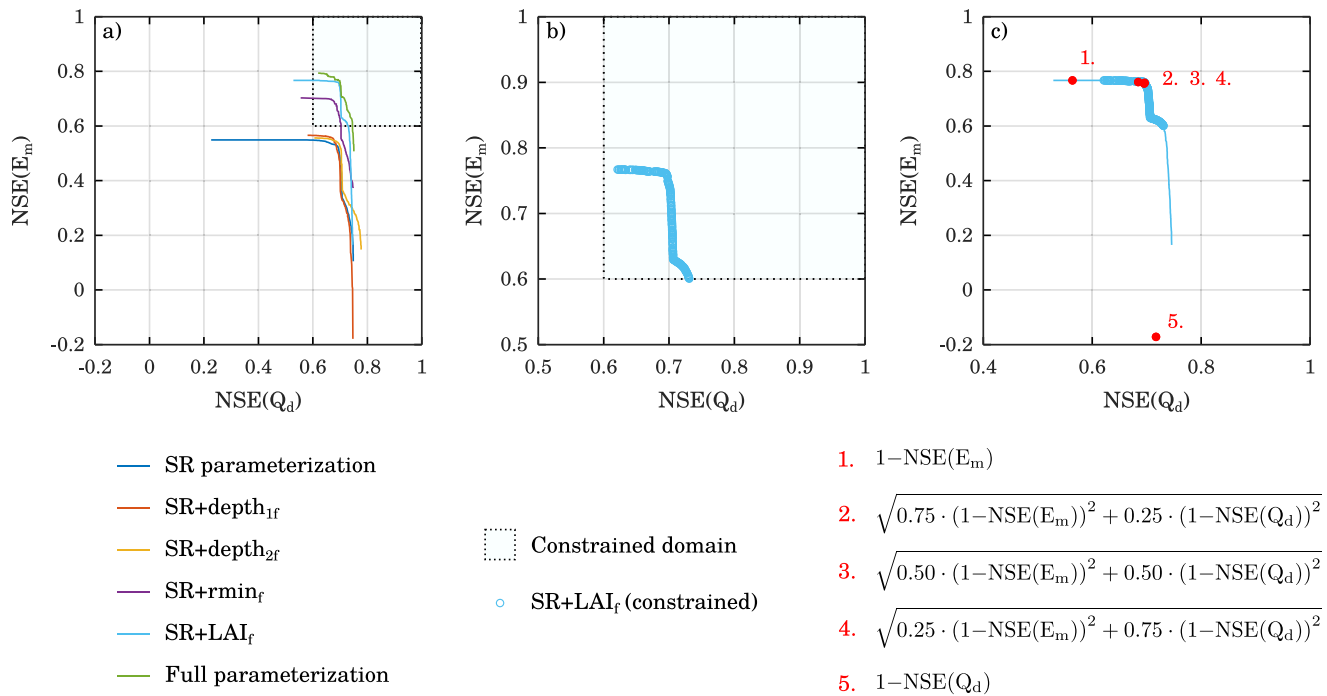

**Figure 5.** Results of the Pareto-based sensitivity analysis and the multiobjective calibration for station R-5055. (a) Pareto fronts corresponding to the six parameterizations evaluated during the Pareto-based sensitivity analysis. (b) Pareto front resulting from the constrained multiobjective calibration for the selected parameterization. (c) Comparative view of the optimal solutions obtained with the Pareto optimization exercises and the five single-objective experiments for the objectives indicated below the panels (see also Table 2).

The widest boxplots in C1 correspond to $D_m$ and $rout_1$, and therefore are the main drivers of the $NSE(Q_d)$ improvement. In C2, $LAI_f$ is the only parameter responsible for the $NSE(E_m)$ changes, suggesting that small changes of $LAI_f$ lead to great improvements in terms of evaporation performance, and corroborating prior results from the Pareto-based sensitivity analysis. Results for C1 and C2 also confirm the sensitivities estimated by DELSA for $D_m$ and $rout_1$ regarding $NSE(Q_d)$ (Figure 4a), and for $LAI_f$ in relation to $NSE(E_m)$ (Figure 4b). The changes of $NSE(Q_d)$ and $NSE(E_m)$ in C3 are, in turn, less evident when seen through the lens of parameter distributions. These three behaviors are also evinced by the abrupt changes of parameter values between the clusters, particularly for $b_l$, $W_S$, $D_m$, $d_2$, and $rout_1$.

Regarding station R-5005, three clusters were defined with similar model behaviors to those described above (Figure S5 in Supporting Information S1). In this case, $b_l$, $d_2$, and $rout_1$ represent the widest boxplots in the region of $NSE(Q_d)$ enhancement, while $rmin_f$ dominates the $NSE(E_m)$ increase. Finally, only one cluster was defined for station R-2011 (Figure S11 in Supporting Information S1), where the small parameter ranges are due to the marginal NSE improvement found in the constrained objective space (Figure S9b in Supporting Information S1).

### 5.4. NSE Decomposition and Single-Objective Experiments

Based on Equation 5, the three components of NSE, $r$, $\alpha$, and $\beta$, were calculated for both $NSE(Q_d)$ and $NSE(E_m)$ and each of the studied catchments, and the results are shown in Figure 8, Figures S6 and S12 in Supporting Information S1. The ranking of the NSE statistics is evident for all the components in the case of stations R-5055 and R-5005, and $r$ is the most contributing factor to it as it manifests an identical NSE ordering to that presented in the Pareto front (see Figure 8b and Figure S6b in Supporting Information S1 and compare to Figure 8a and Figure S6a in Supporting Information S1). Remarkably, cluster C3 (Figure 7) is also noticeable in the NSE decomposition for R-5055 (Figure 8), providing evidence of a complex and nonconvex objective space. The small NSE improvement evinced for R-2011 (Figure S12 in Supporting Information S1) is, however, a symptom of a weak trade-off that is also reflected in the short ranges of $r$, $\alpha$, and $\beta$.

$\alpha$ values are concentrated in the top-left quadrant of Figure 8c, Figures S6c and S12c in Supporting Information S1, revealing a generalized underestimation of the variability for the streamflow ($\alpha_Q < 1$) and an overestimation of the variability for evaporation ($\alpha_E > 1$). As expected, $\beta$ values follow a straight line with negative slope (Equation 7), and its magnitude is close to the value of $Q_{ref}/E_{ref}$ computed for each catchment (Figure 8d, Figures S6d and S12d in Supporting Information S1). Lines are mostly located in the bottom-right quadrant, suggesting an overall overestimation of the mean streamflow ($\beta_Q > 1$) and an underestimation of the mean evaporation ($\beta_E < 1$).

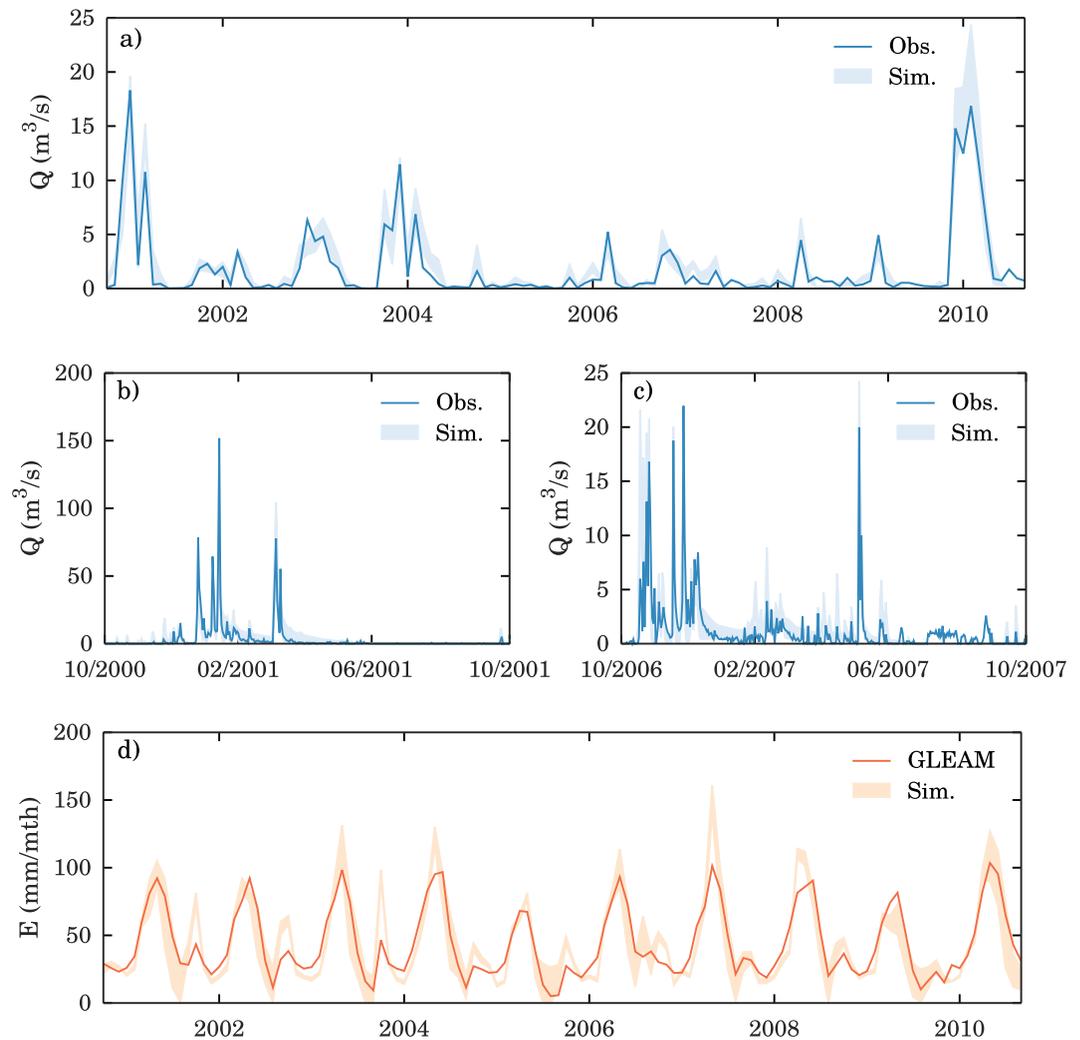

**Figure 6.** Predictive intervals corresponding to the multiobjective calibration for station R-5055. (a) Monthly streamflow simulations and observations for the complete study period. (b and c) Daily streamflow simulations and observations for two different subperiods with a marked contrast in the magnitude of streamflow. (d) Monthly evaporation simulations and Global Land Evaporation Amsterdam Model (GLEAM) data for the complete study period.

Lastly, the single-objective calibration experiments for R-5055 tested the ability to attain Pareto-optimal solutions using SCE-UA (Figures 5c and 8a). Although SCE-UA solutions mostly reach the Pareto front and resemble its shape, some important features of the objective space are not well reproduced. Notably, the point corresponding to the streamflow-only calibration yielded a suboptimal solution in comparison with the multiobjective experiment (Figures 5c and 8a). Both calibration strategies will be discussed and further analyzed in next section.

## 6. Discussion

### 6.1. Integrating Streamflow and Evaporation Data Into Sensitivity Analysis

The profusion of satellite-based and model evaporation data has proven to be a promising step toward a better understanding of the hydrologic system. Many studies have included evaporation as an evaluation variable to gain further insight and strengthen the results for the streamflow-only calibration (e.g., Bouaziz et al., 2021; Rakovec et al., 2019; Yeste et al., 2020). Calibration efforts including evaporation have been normally carried out for multiple hydrologic variables either through aggregating functions (Dembélé, Ceperley, et al., 2020; Dembélé, Hrachowitz, et al., 2020; Demirel et al., 2018; Széles et al., 2020) or by following pure Pareto approaches (Koppa et al., 2019; Nijzink et al., 2018), and outline the potential for enhancing the representation of hydrologic processes within the modeling framework.

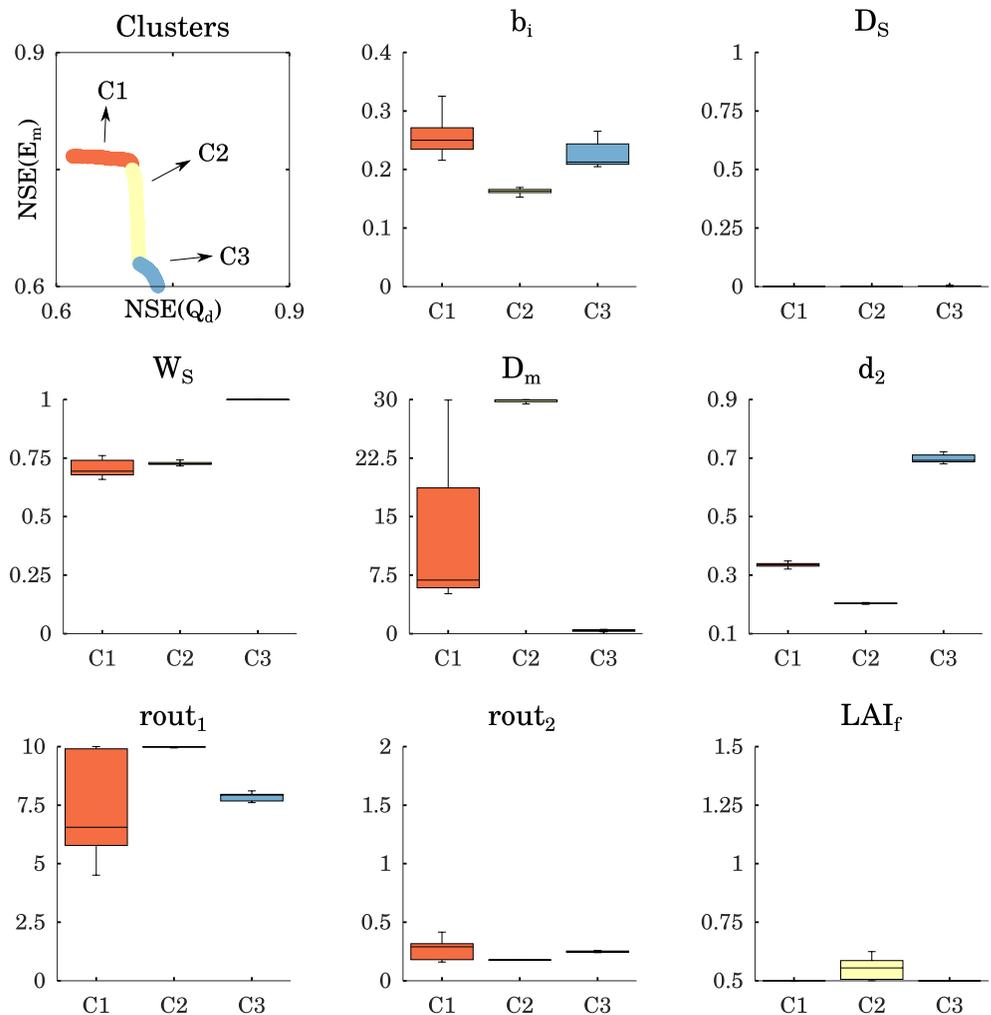

**Figure 7.** Parameter distributions from the multiobjective calibration exercise for station R-5055. The Pareto front was divided into three clusters representing well-differentiated regions of model performance: C1 corresponds to the highest NSE($E_m$) estimates; C2 contains solutions where NSE($Q_d$) remains constant and high and NSE($E_m$) improves; C3 is in the region with the highest NSE($Q_d$) values and lowest NSE($E_m$) estimates. The parameter distributions were evaluated for the three clusters.

Despite the increasing number of works including evaporation as part of the analysis, there is still no consensus on how evaporation data can be used to improve hydrologic models (Dembélé, Hrachowitz, et al., 2020). Moreover, the evaporation-related parameters are rarely considered during the sensitivity analysis stage, where streamflow has traditionally played a major role. In the case of the VIC model, this has resulted in numerous parameters with unknown sensitivities (Sepúlveda et al., 2022).

In this respect, and depending on the studied catchment, up to 5 vegetation parameters out of the 12 vegetation parameters initially selected were identified as important to NSE($E_m$) with DELSA, although their effect was almost unnoticeable for NSE($Q_d$) (Figure 4, Figures S2 and S8 in Supporting Information S1). Particularly, $LAI_f$ and $rmin_f$ were among the most important parameters in terms of evaporation performance for all the catchments, which is consistent with the findings of Sepúlveda et al. (2022). Likewise, their effect was found to be negligible for NSE($Q_d$), though it is expected to become important at annual scale according to Melsen and Guse (2019). For their part, the soil and routing parameters reflected similar sensitivities to those reported in previous studies (e.g.,

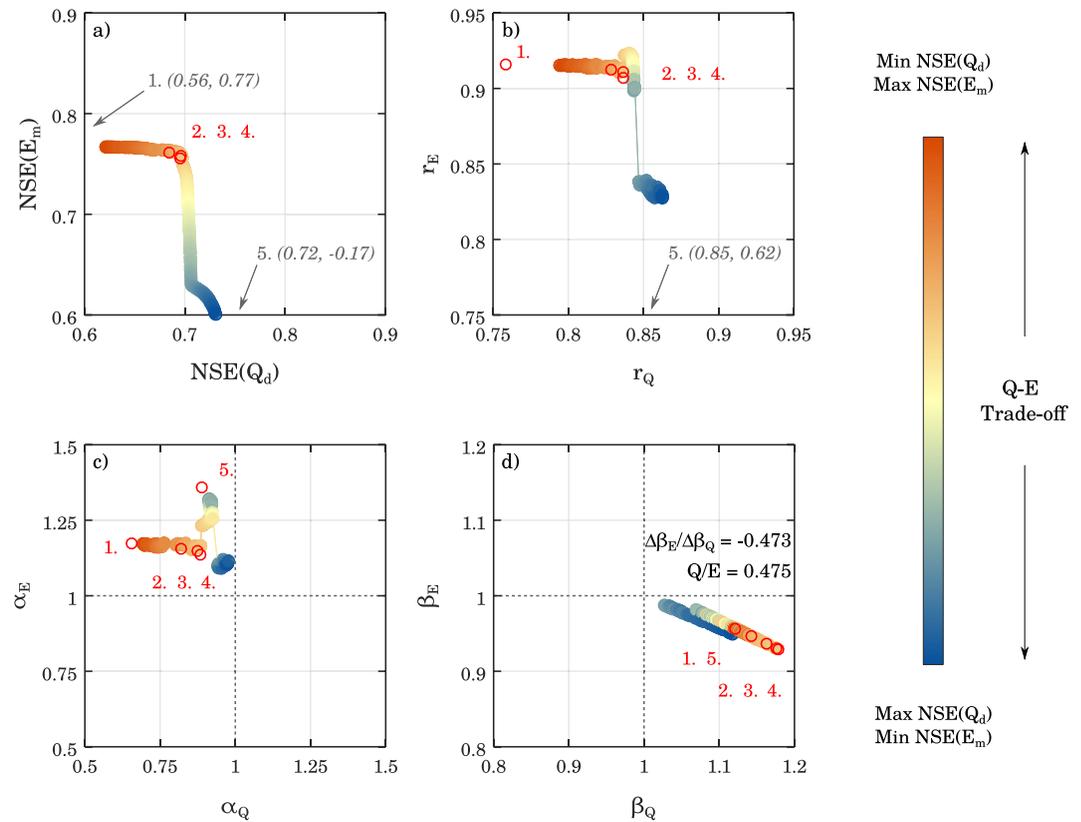

**Figure 8.** Nash-Sutcliffe Efficiency (NSE) decomposition into $r$, $\alpha$, and $\beta$ and trade-off evaluation for station R-5055. The NSE ordering is first determined for the Pareto front corresponding to the multiobjective calibration using a divergent color scale (a). The NSE ordering is then superimposed on the decomposition in $r$, $\alpha$, and $\beta$ to evaluate the extent of the $Q$-$E$ trade-off (b–d). Equation 7 was checked for the $\beta$ components of NSE($Q_d$) and NSE($E_m$), and the resulting values are indicated in panel d. The results of the NSE decomposition for the solutions of the five single-objective experiments are depicted in every panel in red, or indicated in gray if outside of the plotting region.

Demaria et al., 2007; Gou et al., 2020; Lilhare et al., 2020; Melsen & Guse, 2019; Melsen et al., 2016; Mendoza et al., 2015; Rakovec et al., 2014).

Although the results of the Pareto-based sensitivity analysis are largely in line with the DELSA first-order sensitivity measures for the studied catchments, the relative importance of each parameterization in terms of joint model performance for both NSE($Q_d$) and NSE($E_m$) could not be uncovered by only using DELSA. In this sense, the full parameterization experiments were still subject to overparameterization as they included all the parameters identified with DELSA. The Pareto-based sensitivity analysis successfully overcame this limitation, and its strength lied in its ability to directly pinpoint the most parsimonious implementation leading to a similar performance to that corresponding to the full parameterization.

Thus, no more than two vegetation parameters were required in all cases: $LAI_f$ for station R-5055 (Figure 5), $LAI_f$ and $rmin_f$ for R-5005 (Figure S3 in Supporting Information S1), and $depth_{1f}$ and $depth_{2f}$ for R-2011 (Figure S9 in Supporting Information S1). As suggested in Sepúlveda et al. (2022), these differences are likely connected to the existing hydroclimatic gradient between the stations: stations R-5055 and R-5005 are located in the Guadalquivir River Basin, the southern-most, and one of the most semiarid basins in Spain (Yeste et al., 2018), while R-2011 belongs to a more humid climate characteristic of the Duero River Basin (Yeste et al., 2020, 2021).

## 6.2. Benchmark Comparison of Calibration Strategies

### 6.2.1. Multiobjective Calibration and Single-Objective Experiments

The optimization approach developed in this work serves the purpose of both a sensitivity analysis and a calibration method, the only difference being the use of constraints during optimization. Whereas the unconstrained implementation included poor estimates for NSE($Q_d$) and NSE($E_m$) that are of little interest from a calibration perspective (Figure 5a, Figures S3a and S9a in Supporting Information S1), the use of constraints during the multiobjective calibration produced solutions within the predefined limits of acceptability (Figure 5b, Figures S3b and S9b in Supporting Information S1).

The results for the single-objective experiments mostly reproduced the shape of the Pareto front and yielded optimal solutions (Figure 5c), although they point out important technical issues concerning local optimality. In particular, the streamflow-only calibration demonstrates that the optimization algorithm became stuck in a local minimum (Figure 5c), probably due to the higher density of points revealed during the Monte Carlo experiment for this area (Figure 3c). Additionally, the calibration results for the weighted distances could barely be differentiated from each other as the solutions tended to cluster together in the region closest to the ideal solution (i.e., top-right corner of the scatterplot, Figure 5c).

A similar approach was carried out in Fowler et al. (2018) in an assessment of the Differential Split Sample Test (DSST; Klemeš, 1986) for two objectives representing the calibration-evaluation trade-off for streamflow through both multiobjective and single-objective experiments. They concluded that none of the single-objective methods could satisfy the limits of acceptability for both objectives at the same time if only the objective for the calibration period was optimized. Nonetheless, single-objective optimization is still a valid alternative to a full Pareto optimization procedure when working with multiple objectives if the objectives are combined in a single aggregating metric such as the weighted Euclidean distance. As shown in Figure 5c, the performance gain for evaporation with the weighted metric experiments was substantial in comparison with the slight performance loss detected for streamflow. Future efforts will be directed at exploring this issue in more detail for a large sample of Spanish catchments.

### 6.2.2. NSE Diagnosis

An important drawback of using NSE as an objective during optimization is that the calibration process is ultimately bounded to the relation of $r$, $\alpha$, and $\beta$ that stems from its very definition. As shown in Gupta et al. (2009), the first derivative of Equation 5 with respect to $\alpha$ indicates that NSE is optimum when $\alpha = r$. This will favor (but not strictly impose) parameter combinations that underestimate $\alpha$ as $r$ is always less than 1, leading to an underestimation of high-flows and an overestimation of low-flows (Gupta et al., 2009; Mizukami et al., 2018). While this is visible for $\alpha_Q$, $\alpha_E$ is always greater than 1 for all the catchments (Figure 8c, Figures S6c and S12c in Supporting Information S1), confirming previous research using VIC and NSE as a performance metric for streamflow and evaporation in Rakovec et al. (2019) and Yeste et al. (2020). A possible explanation for this is that $r$ values are generally higher for NSE($E_m$) than for NSE($Q_d$). On the other hand, Equation 7 reveals that $\beta$ values are subject to a linear trade-off when the steady-state is achieved that is exclusively dependent on reference data for streamflow and evaporation. $\beta$ lines in Figure 8d, Figures S6d and S12d in Supporting Information S1 indicate that the steady-state is always reached by overestimating $Q$ and underestimating $E$ in all cases, which is also in agreement with the findings of Rakovec et al. (2019) and Yeste et al. (2020).

It has to be recognized that similar limitations are inherent to any aggregating metric one may choose. For example, the KGE (Gupta et al., 2009), which is together with NSE the most used performance metric in hydrologic studies (Clark et al., 2021; Knoben et al., 2019), is based on a Euclidean distance formulation for $r$, $\alpha$, and $\beta$. Limitations for $\beta$ shown in Equation 7 apply to KGE as well by definition. It is also known that $\beta$ values tend to be overestimated with KGE if the mean value of reference data is small (Clark et al., 2021; Santos et al., 2018). This constitutes a potential pitfall for regions including arid conditions such as the Iberian Peninsula (García-Valdecasas Ojeda et al., 2020, 2021; Páscoa et al., 2017; Vicente-Serrano et al., 2014), where Rakovec, Kumar, Mai, et al. (2016) observed poor estimates of KGE in a large-sample evaluation of streamflow performance for 400 European river basins.

Besides its intrinsic limitations, the decomposition of NSE into $r$, $\alpha$, and $\beta$ provides valuable information as to why NSE values are as they are, and can help elucidate possible causes of model failure as well as potential avenues for improving model performance. The simultaneous consideration of NSE($Q_d$) and NSE($E_m$) is also a necessary step in order to explore model performance and alleviate the shortcomings of using a single system-scale metric (Clark et al., 2021), which has been an ever-present goal in this study.

## 7. Conclusions

We present a framework to incorporate streamflow and evaporation data into the sensitivity analysis and calibration stages of the hydrologic modeling exercise. The framework is articulated into four consecutive steps inspired by the definition of the MOOP in order to study the streamflow and evaporation trade-offs, giving rise to what we have referred to as "Pareto-based sensitivity analysis" and "multiobjective calibration." The proposed approach was then applied to three Spanish catchments using VIC.

The overall performance for streamflow and evaporation was first explored in a Monte Carlo experiment using daily and monthly estimates of NSE. As a result, the daily NSE(Q) and the monthly NSE(E) were selected as the metrics of reference. A parameter screening procedure based on the DELSA method was then conducted to study parameter sensitivities, and the most influential parameters were chosen in each case. This selection was polished during the Pareto-based sensitivity analysis, and permitted assessing the gains/losses in performance and identifying the most parsimonious parameterization that produced a proper approximation to both reference streamflow and evaporation. The multiobjective calibration was performed for the selected parameterization using constraints applied to the objective space to avoid poor performance estimates, and the solutions were compared to five single-objective optimization experiments for one of the catchments. The multiobjective calibration outperformed the single-objective experiments and produced a better representation of the streamflow and evaporation trade-offs, suggesting the possibility to obtain satisfactory estimates for both variables with hardly any performance loss for streamflow compared to the VIC calibration only against streamflow data.

The use of multiple-criteria methods as in this study is a step forward in the direction of "What is my model good for?" rather than "How good is my model?" (Clark et al., 2021), and has guided the development of the approach and the experimental design since their early inception. Answers to the former question started coming during sensitivity analysis, with DELSA and Pareto-based, and became clearly visible after calibrating VIC. As shown in this study, multiobjective optimization approaches can play the role of sensitivity analysis methods to evaluate parameter sensitivities by gradually increasing model complexity, and can help determine optimal model performance using constraints applied to the objective space. Our results will contribute to gain further insight into the nature of the trade-off resulting from the simultaneous adjustment to streamflow and evaporation data.

## Data Availability Statement

Computer code for NSGA-II (Deb et al., 2002) is provided in pymoo (Blank & Deb, 2020). Computer code for MOFPA (Yang et al., 2014) is written in Matlab/GNU Octave and is available online (https://www.mathworks.com/matlabcentral/fileexchange/74750-multi-objective-flower-pollination-algorithm-mofpa/). Precipitation and temperature data used in this work are available from Serrano-Notivoli et al. (2017) and Serrano-Notivoli et al. (2019). Streamflow observations are provided by CEDEX and are available online (https://ceh.cedex.es). Evaporation data have been collected from GLEAM (Martens et al., 2017; Miralles et al., 2011). Basemap in Figure 2 has been made with OpenStreetMap (https://www.openstreetmap.org/copyright).

**Acknowledgments**
All the simulations were conducted in the ALHAMBRA cluster (http://alhambra.ugr.es) of the University of Granada. This work was funded by the project P20_00035, funded by the FEDER/Junta de Andalucía-Consejería de Transformación Económica, Industria, Conocimiento y Universidades, the project CGL2017-89836-R funded by the Spanish Ministry of Economy and Competitiveness, with additional support from the European Community Funds (FEDER), the project PID2021-126401OB-I00, funded by MCIN/AEI/10.13039/501100011033/FEDER Una manera de hacer Europa and the project LifeWatch-2019-10-UGR-01 funded by FEDER/Ministerio de Ciencia e Innovación. The first author was supported by the Ministry of Education, Culture and Sport of Spain through an FPU Grant (reference FPU17/02098) and an Aid for Research Stays in the Hydrology and Quantitative Water Management Group of Wageningen University (reference ESTI9/00169). Funding for open access charge was provided by Universidad de Granada/CBUA.